\documentclass[twoside]{ilcws10}
\usepackage[latin1]{inputenc}
\usepackage[dvips]{graphicx,epsfig,color}
\usepackage{wrapfig,rotating}
\usepackage{amssymb,amsmath,array}

\begin{document}
\title{Simulation Study of FPCCD Vertex Detector} 
\author{Kohei Yoshida$^1$, Daisuke Kamai$^1$, Akiya Miyamoto$^2$, Yasuhiro Sugimoto$^2$, \\
Yosuke Takubo$^1$, and Hitoshi Yamamoto$^1$
\vspace{.3cm}\\
1- Department of Physics, Tohoku University, Sendai , Japan
\vspace{.1cm}\\
2- High Energy Accelerator Research Organization (KEK), Tsukuba, Japan \\
}

\maketitle

\begin{abstract}
We are developing the vertex detector with a fine pixel CCD (FPCCD) for the international linear collider (ILC), whose pixel size is $5 \times 5$ $\mu$m$^{2}$. To evaluate the performance of the FPCCD vertex detector and optimize its design, development of the software dedicated for the FPCCD is necessary. We, therefore, started to develop the software for FPCCD. In this article, the status of the study is reported.
\end{abstract}

\section{Introduction}
Measurement of the Higgs branching ratio and Higgs self-coupling is the most important program at ILC to investigate the electroweak symmetry breaking. For that purpose, it is necessary to identify the quark flavor of the jets to select the Higgs decay event and reject background effectively. The excellent flavor tagging of the jets is, therefore, indispensable at ILC.

The precise measurement of the decay vertexes and the secondary tracks is important for the flavor tagging. Since the vertex detector is located at the nearest position from the interaction point (IP), performance of the flavor tagging depends on that of the vertex detector. The impact parameter resolution of $5 \oplus 10/(p\beta \sin^{3/2} \theta)$ $\mu$m has been set as the goal of the vertex detector performance \cite{oregon}.

\begin{wrapfigure}{r}{0.42\columnwidth}
\centerline{
\includegraphics[width=0.4\columnwidth]{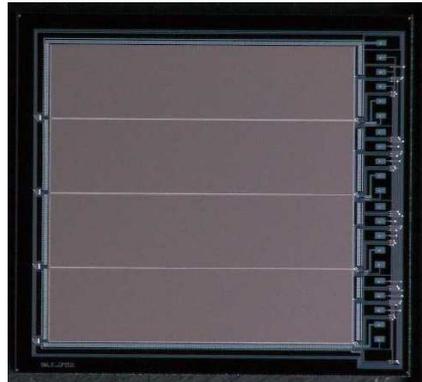}}
\caption{A picture of the FPCCD prototype.}\label{Fig:fpccd}
\end{wrapfigure}

In addition to the position resolution, the hit occupancy is a big problem for the vertex detector at ILC. In the beam crossing, the large number of low energy electron-positron pairs, called pair background, will be generated at IP. The pair background makes many hits in the vertex detector, and the hit occupancy especially at the first layer of the vertex detector becomes large. If all the hits for one beam train are accumulated, it was estimated that the hit occupancy will become $\sim 10$\% for the pixel size of $25 \times 25$ $\mu$m$^{2}$. We should suppress the hit occupancy below 1\% for the reasonable track resolution.

To solve the problem of the hit occupancy, a FPCCD (Fine PixelCCD) has been proposed, whose pixel size is $5 \times 5$ $\mu$m$^{2}$. Since the pixel size of the FPCCD is very small, it has good position resolution and reduces the hit occupancy. The first prototype was developed in 2008 \cite{fpccd}, whose pixel size is $12 \times 12$ $\mu$m$^{2}$ as shown in Fig. \ref{Fig:fpccd}. To evaluate the FPCCD performance in the ILC experiment and use its results for the design optimization, development of the software dedicated for the FPCCD is necessary. We, therefore, started to develop the software for FPCCD and studied the pixel occupancy. 

\begin{wrapfigure}{r}{0.5\columnwidth}
\centerline{\includegraphics[width=0.45\columnwidth]{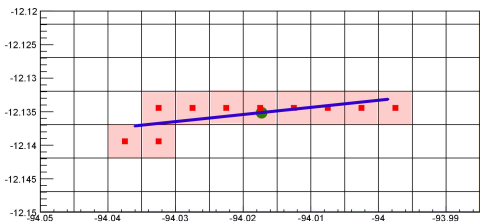}}
\caption{The pixel hits in a FPCCD chip obtained by the digitizer for a single electron with 1 GeV. The green point is the hit position at the center of the sensor chip for a particle penetrating the vertex layers by Mokka output. The blue line shows the expected track and red points are the center position of the pixel hits.}
\label{Fig:digitized}
\end{wrapfigure}

\section{Development of FPCCD digitizer}
At first, we develop the FPCCD digitizer, which makes pixel hits from the hit information obtained by the simulation. Mokka \cite{mokka} in the ILC Soft v07-02 \cite{ilcsoft} is used for the full simulation in the detector, where ILD\_00fwp01 is used as the detector model \cite{ild}. The anti-DID magnetic field is taken into account in the simulation. The vertex detector of ILD\_00fwp01 consists of three double layers. The double layer has two sensor chips attached in both sides of the layer. 

The information of the hit position and momentum at the center of the sensor chip for a particle penetrating the vertex layers is included in the Mokka output. In the FPCCD digitizer, the pixels to be fired by the particle are calculated by using the information. The algorism of the procedure is as follows. The particle track in the vertex chip is calculated by using the hit information obtained by Mokka. At that time, the track is approximated by a linear track. Then, it is transformed to the track at the local coordinate on each sensor chip which is shown as blue line in Fig. \ref{Fig:digitized}. The fired pixels are identified by using the track information, where the threshold of the path length in each pixel is not set in this study. The pixel hits for a single electron with 1 GeV are shown as the red points in Fig. \ref{Fig:digitized}, which are obtained by the FPCCD digitizer. We can see that the pixel hits are obtained correctly along the track on the local coordinate of the sensor chip.

\begin{table}
\centerline{\begin{tabular}{|l|r|r|r|}
\hline
& RDR-Nominal & RDR-LowP & SB2009wTF \\\hline
\# of bunches & 2625 & 1320 & 1312 \\ \hline
$\sigma_{x}$ (nm) & 639 & 474 & 474 \\ \hline
$\sigma_{y}$ (nm) &  5.7 & 3.8 & 3.8 \\ \hline
$L$ ($10^{34}$ cm$^{-2}$s$^{-1}$) & 2.0 & 2.0 & 2.0 \\ \hline
\end{tabular}}
\caption{Beam parameters of RDR-Nominal, RDR-LowP, and SB2009wTF for $\sqrt{s} = 500$ GeV.}
\label{tb:beam}
\end{table}

\section{Estimation of pixel occupancy}
The pixel occupancy is evaluated by using the FPCCD digitizer. The electron-positron pair backgrounds are created by GUINEA-PIG. The hit occupancy is evaluated for the beam parameters, RDR-Nominal, RDR-LowP, and SB2009wTF with $\sqrt{s} = 500$ GeV. The characteristics of each beam parameter are as shown in Table \ref{tb:beam}. The MC statistics of the pair background corresponds to that of 169, 68, and 59 beam-bunch crossing (BX) for RDR-Nominal, RDR-LowP, and SB2009wTF, respectively.

At the innermost layer, we have 62, 177, and 164 hits/cm$^2\cdot$BX for RDR-Nominal, RDR-LowP, and SB2009wTF, respectively. Those results correspond to the pixel occupancy for one train of 4.1\%, 5.8\%, and 5.4\% for RDR-Nominal, RDR-LowP, and SB2009wTF, respectively, as shown in Table \ref{tb:occupancy}.

\begin{wrapfigure}{r}{0.6\columnwidth}
\centerline{\includegraphics[width=0.57\columnwidth]{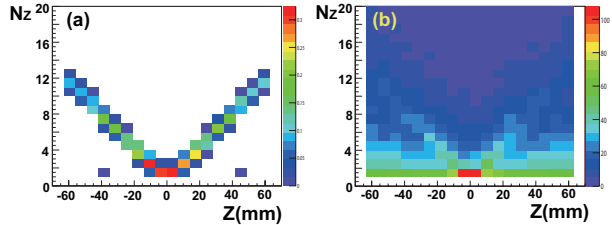}}
\caption{The number of the pixel hits in the $Z$ direction ($N_{Z}$) v.s. Z position for 1 GeV electrons (a) and pair backgrounds (b) at the innermost sensor chip.}\label{Fig:nz}
\end{wrapfigure}


Since the pixel size of the FPCCD is very small, it is possible to reject the background hits by using shape of the hit clusters. We, therefore, use the width of the hit cluster along $Z$ and $\phi$ directions, where $Z$ is the beam axis and $\phi$ is the azimuth angle. To compare the distribution of the pixel his with pair backgrounds, the single electrons with the energy of 50, 100, and 500 MeV and 1GeV are used as the signal tracks. At this time, they are emitted at IP with the uniform polar and azimuth angle.

The single electrons with the energy of 1 GeV make the pixel hits proportional to the polar angle as shown in Fig. \ref{Fig:nz}(a). Since most of the pair backgrounds have energy less than 100 MeV, the polar angle of a particle hitting the vertex layers are larger than the signal. For that reason, the pair backgrounds make a few pixel hits in the $Z$ direction and penetrate the same sensor chip several times. The hit distribution becomes uniform in $Z$ direction as shown in Fig. \ref{Fig:nz}(b). We require the number of the pixel hits in the $Z$ direction ($N_{Z}$) to be $3Z/R< N_{Z} <3Z/R + 2$ to reject the pair background, where $R$ is radius of each layer from IP.

\begin{wrapfigure}{r}{0.6\columnwidth}
\centerline{\includegraphics[width=0.57\columnwidth]{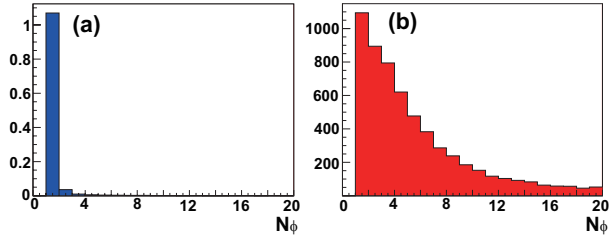}}
\caption{The number of the pixel hits in the $\phi$ direction ($N_{\phi}$) for 1 GeV electrons (a) and pair backgrounds (b) at the innermost sensor chip.}\label{Fig:nphi}
\end{wrapfigure}

In the plane perpendicular to the beam axis, the single electrons with the energy of 1 GeV go through the sensor chips and makes a few pixel hits in the $\phi$ direction because the track of the particle is almost straight. We, therefore, have the small number of the pixel hits as shown in Fig. \ref{Fig:nphi}(a). On the other hand, pair background tracks have shallow incident angle to the sensor chips because of their low transverse momentum. For that reason, background hits many pixels in the $\phi$ direction as shown in Fig. \ref{Fig:nphi}(b). We selected the number of pixel hits in the $\phi$ direction ($N_{\phi}$) to be $N_{\phi} \leq 2$.

The effective pixel occupancy is investigated after the selection cuts as shown in Table \ref{tb:occupancy}, and it can be reduced to be below 1\%. It is, however, optimized only for a particle with the energy of 1 GeV. For that reason, the selection efficiency for particles with low energy below 1 GeV becomes worse than the high energy tracks as shown in Table \ref{tb:eff}. The selection cut should be investigated to have better efficiency for the low energy tracks.

\begin{table}
\begin{center}
\begin{tabular}{|l|c|c|c|c|} \hline 
& \multicolumn{2}{|c|}{Before cut} & \multicolumn{2}{|c|}{After cut} \\ \hline
& 1a & 1b & 1a & 1b \\ \hline 
RDR-Nominal & 4.07\% & 2.33\% & 0.13\% & 0.08\% \\ \hline 
RDR-LowP & 5.79\% & 3.05\% & 0.17\% & 0.11\% \\ \hline 
SB2009wTF & 5.39\% & 3.02\% & 0.17\% & 0.11\% \\ 
\hline 
\end{tabular}
\caption{The effective occupancy before and after selection cut for the inner (a) and outer (b) chips in the first layer.}
\label{tb:occupancy}
\end{center}
\end{table}

\begin{table}
\begin{center}
\begin{tabular}{|c|c|c|}
\hline  & 1a & 1b \\ 
\hline 50MeV & 46.99\% & 46.61\% \\ 
\hline 100MeV & 89.82\% & 81.63\% \\ 
\hline 500MeV & 96.22\% & 97.45\% \\ 
\hline 1GeV & 97.98\% & 99.42\% \\ 
\hline 
\end{tabular}
\caption{Selection efficiency of electrons with particular energy for the inner (a) and outer (b) chips in the first layer.}
\label{tb:eff}
\end{center}
\end{table}

\section{Conclusions}
We are developing the vertex detector with FPCCD for ILC. To evaluate the performance of the FPCCD vertex detector and optimize its design, we developed the software for FPCCD and studied the pixel occupancy. At first, we develop the FPCCD digitizer, which makes pixel hits from the hit information obtained by the simulation. By using the FPCCD digitizer, the pixel occupancy by the pair background is evaluated. Without any treatment, the pixel occupancy is around 5\% at the innermost chip of the vertex detector. The filter of background hits based on the width of the hit cluster is developed. Optimizing the filter for tracks with energy greater than 1 GeV, the hit occupancy can be reduced below 1\%.

\section{Acknowledgments}
The authors would like to thank all the members of the ILC physics subgroup \cite{subgroup} for useful discussions. This work is supported in part by the Creative Scientific Research Grant (No. 18GS0202) of the Japan Society for Promotion of Science and the JSPS Core University Program.


\begin{footnotesize}


\end{footnotesize}


\end{document}